\documentstyle{mn}                 %

\def\Ha{H$\alpha$}

\input{psfig.tex}
\title[Supernova 1994aj]
{Supernova 1994aj: a probe for pre-supernova evolution 
and mass loss from the progenitor.
\thanks{Based on observations collected at ESO-La Silla (Chile)}}

\author[Benetti et al.]
{S. Benetti$^1$, E. Cappellaro$^2$, I.J. Danziger$^{3,4}$,
M. Turatto$^{1,2}$, F. Patat$^{3,5}$, \and and M. Della Valle$^5$
\\
$^1$European Southern Observatory, Alonso de Cordova 3107, Vitacura,
Casilla 19001 Santiago 19, Chile\\ 
$^2$Osservatorio Astronomico di Padova, vicolo dell'Osservatorio 5,
I-35122 Padova, Italy\\
$^3$European Southern Observatory, Karl-Schwarzschild-Strasse 2,
D-8046 Garching bei M\"unchen, Germany\\
$^4$Osservatorio Astronomico di Trieste, via G.B. Tiepolo 11, I-34131
Trieste, Italy\\
$^5$Dipartimento di Astronomia, Universit\'a di Padova, vicolo
dell'Osservatorio 5,I-35122 Padova, Italy\\
}

\date{Received ................; accepted ................}

\begin{document}

\maketitle

\begin{abstract}

Extensive photometric and spectroscopic observations of SN~1994aj
until 540d after maximum light have been obtained. The photometry
around maximum suggests that the SN belongs to the Type II Linear
class, with a peak absolute magnitude of $M_{\rm V} \sim -17.8$
(assuming $H_{0}=75 km~s^{-1}~Mpc^{-1}$). The spectra of SN~1994aj
were unusual with the presence of a narrow line with a P-Cygni profile
on the top of the broad Balmer line emission. This narrow feature is
attributed to the presence of a dense superwind surrounding the SN. At
100-120 days after maximum light the SN ejecta starts to interact with
this CSM. The SN luminosity decline rates slowed down (${\gamma}_{\rm
R}=0.46$ \,mag~(100d)$^{-1}$), becoming less steep than the average
late luminosity decline of normal SNII ($\sim 1$
\,mag~(100d)$^{-1}$). This dense ($\dot{M}/u_{\rm w} \sim 10^{15}$
g/cm) wind was confined to a short distance from the progenitor
($R_{\rm out} = \sim 5\times 10^{16}$ cm), and results from a very strong mass
loss episode ($\dot{M} = 10^{-3} \,M_{\odot}/yr$), which terminated
shortly before explosion ($\sim 5-10$ yr).

\end{abstract}


\section{Introduction}

The massive progenitors of SN~II experience, during their life times,
strong mass loss.  In different phases of its evolution an $8 - 15
M_{\odot}$ star, a typical range for the initial mass of a SN~II
progenitor, may eject several solar masses of material with different
velocities and mass loss rates. Therefore in the neighbourhood of Type II
SNe dense circumstellar material (CSM) may be expected to be distributed
according to the detailed mass loss history of the progenitor.\\
In many cases there is not observational evidence of this material,
whereas in others this material determines its observed photometric and
spectroscopic properties \cite{tur}.

In particular a subclass of SN~II has been identified showing narrow
emission lines in the spectrum (e.g.  SN~1988Z; Turatto et al., 1993,
1997), and therefore called IIn \cite{schl}. Probably, these are
normal type II explosions occurring in a dense CSM which is confined
very close to the SN so that the ejecta-CSM interaction began shortly
after explosion \cite{cd}.

In addition, a number of SN~II, SN 1970G (Fesen 1993), 1979C (Fesen \&
Matonick 1993), 1980K (Leibundgut 1991, Fesen \& Matonick 1994)
and 1986E (Cappellaro et al. 1995a), all of the Linear subclass, have
been recovered a few years after explosion at a luminosity greatly
exceeding that expected from the normal radioactive decay and again
explained by enhanced ejecta-CSM interaction beginning 1-2 years after
explosion.\\
An outstanding and so far unique case was that of SN~1984E whose
spectral features showed an unusual profile with a narrow P-Cygni
component on top of the usual broad feature. The narrow feature was
attributed to a superwind with a velocity of $\sim 3000$
km\,sec$^{-1}$
\cite{dop}. Unfortunately only a couple of spectra near maximum were
obtained for this SN.

Here we present and discuss the case of SN~1994aj which shows at
early phases the double P-Cygni line profile similar to SN~1984E, and
at late phases the enhanced H$\alpha$ luminosity similar to other SN~IIL.

\section{Observations} \label{obs}

\begin{figure}
\psfig{figure=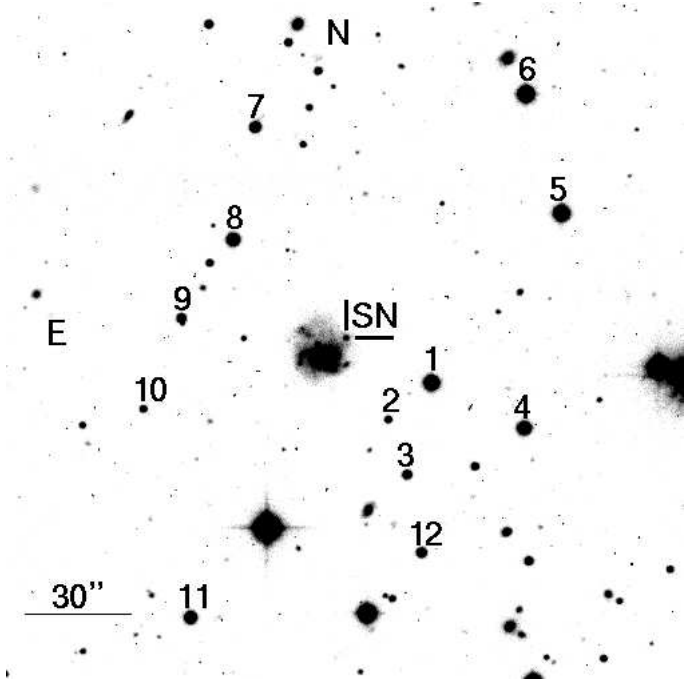,width=9cm,height=9cm}
\caption{SN 1994aj in Anon0903-10 and reference stars. The image is an
R frame taken at D1.54m telescope at La Silla on Apr. 20, 1995.} \label{sn}
\end{figure}

The SN~1994aj was found by Pollas \shortcite{pol} on two films obtained on
Dec. 12.16 and 13.14 UT respectively in a late type ($\sim$~Sc)
anonymous galaxy, at R.A. = 9h03m46s.12, Decl. = -10$^o$27'56".2 (equinox
1950.0). The SN was located 5''. 2W, 5''. 1N of the galaxy nucleus,
superimposed on an external spiral arm (Fig. \ref{sn}).
The first spectrum, obtained several days after discovery \cite{ben1},
indicated that the SN was of type II because of the presence of
hydrogen Balmer lines which showed an unusual double peak profile. As
with SN~1984E a narrow P-Cyg profile appeared to be superimposed on a
much broader component.

This prompted an intense observational effort which, despite the relative
faintness of the SN, about 18 mag at maximum, allowed photometric and
spectroscopic monitoring for over 1 year.

\subsection{Photometry} \label{phot}

V and R-band photometry of SN~1994aj were obtained at ESO-La Silla on
22 nights using five different telescopes. Photometric nights were
used to calibrate, through observations of photometric standard stars
\cite{land}, a sequence of stars around the SN. In turn, the local
sequence was used to calibrate the observations obtained during
non-photometric nights. The magnitudes of the local standards, labeled
in Fig. \ref{sn}, are reported in Tab. \ref{seq}.

\begin{table}
\caption{Magnitudes of the stars of the local sequence identified in
Fig.~1} \label{seq}
\begin{tabular}{ccc}
\hline
~~star~~ & V & R \\
\hline	 
1        & $16.97\pm 0.02$ & $16.71\pm 0.03$ \\
2        & $20.36\pm 0.01$ & $20.03\pm 0.05$ \\
3        & $19.56\pm 0.03$ & $18.92\pm 0.03$ \\
4        & $18.33\pm 0.05$ & $17.21\pm 0.06$ \\
5        & $16.97\pm 0.03$ & $16.49\pm 0.04$ \\
6        & $16.55\pm 0.02$ & $16.23\pm 0.03$ \\
7        & $19.00\pm 0.04$ & $18.08\pm 0.05$ \\
8        & $18.03\pm 0.04$ & $17.39\pm 0.04$ \\
9        & $20.07\pm 0.04$ & $18.94\pm 0.04$ \\
10       & $20.37\pm 0.04$ & $20.01\pm 0.04$ \\
11       & $17.98\pm 0.02$ & $17.64\pm 0.03$ \\
12       & $19.42\pm 0.06$ & $18.56\pm 0.05$ \\
\hline
\end{tabular}
\end{table}

The SN magnitudes have been measured by point spread
function fitting using the Romafot package in MIDAS. This technique,
allowing a good subtraction of the background gives reliable
results even when the SN fades. The resulting magnitudes with the
estimate of the internal errors are reported in Tab.~\ref{obs_tab}.

\begin{table}
\caption{Photometric measurements for SN~1994aj}\label{obs_tab}
\begin{tabular}{lcccl}
\hline
    date     &  J.D.    &        V       &        R        &   instr.    \\
             & 2400000+ &                &                 &              \\
\hline	      	       
    31/12/94 & 49717.7  & $19.46\pm 0.06$ & $19.09\pm 0.04$ &    2.2       \\
      3/1/95 & 49720.8  &                 & $19.19\pm 0.05$ &    3.6       \\
      8/1/95 & 49725.7  & $19.86\pm 0.05$ & $19.39\pm 0.05$ &    3.6       \\
     10/1/95 & 49727.7  & $19.93\pm 0.04$ & $19.51\pm 0.04$ &    NTT+S     \\
     13/1/95 & 49730.8  & $20.09\pm 0.05$ & $19.59\pm 0.05$ &    NTT+S     \\
     15/1/95 & 49732.8  &                 & $19.62\pm 0.03$ &    NTT+E     \\
     16/1/95 & 49733.8  &                 & $19.69\pm 0.04$ &    NTT+E     \\
     28/1/95 & 49745.7  & $20.80\pm 0.06$ & $20.12\pm 0.04$ &    Dutch     \\
     30/1/95 & 49747.7  &                 & $20.28\pm 0.04$ &    NTT+E     \\
      4/2/95 & 49752.7  & $21.55\pm 0.07$ & $20.61\pm 0.05$ &    2.2       \\
     20/2/95 & 49768.7  & $21.91\pm 0.07$ & $21.01\pm 0.06$ &    D1.54     \\
     28/2/95 & 49776.6  &                 & $21.07\pm 0.06$ &    NTT+E     \\
      2/3/95 & 49779.6  &                 & $21.11\pm 0.06$ &    NTT+E     \\
      6/3/95 & 49782.6  & $21.97\pm 0.10$ & $21.09\pm 0.09$ &    Dutch     \\
     29/3/95 & 49805.5  &                 & $21.30\pm 0.06$ &    3.6       \\
     20/4/95 & 49827.6  & $22.50\pm 0.20$ & $21.41\pm 0.09$ &    Danish    \\
     29/4/95 & 49837.5  & $22.57\pm 0.15$ & $21.46\pm 0.06$ &    3.6       \\
 29/5/95$^*$ & 49866.5  & $22.62\pm 0.20$ & $21.50\pm 0.06$ &    NTT+E     \\
    14/11/95 & 50035.3  &                 & $22.21\pm 0.30$ &    NTT+E     \\
    17/03/96 & 50159.6  &                 & $23.02\pm 0.20$ &    3.6       \\
    10/05/96 & 50214.5  &                 & $23.04\pm 0.30$ &    NTT+E     \\
    11/05/96 & 50215.5  &                 & $23.13\pm 0.30$ &    NTT+E     \\
\hline
\end{tabular}
$*$ for this epoch is available also an estimate $B\simeq 23.1$\\
\\
2.2 = ESO/MPI 2.2m telescope + EFOSC2 \\
3.6 = ESO 3.6m telescope + EFOSC1\\
NTT+S = ESO NTT + SUSI\\
NTT+E = ESO NTT + EMMI\\
Dutch = Dutch 0.90m + CCD Camera \\
Danish = ESO/Danish 1.5m telescope + CCD camera\\

\end{table}

\begin{figure}
\psfig{figure=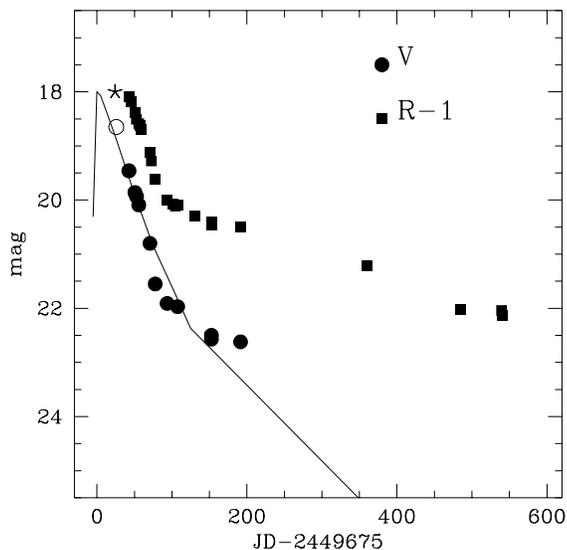,width=8.5cm}
\caption{V and R light curves of SN 1994aj. The open symbol is
a V magnitude from Suntzeff (1994) whereas the starred symbol is the
discovery magnitude by Pollas (1994) . The best fit with the template
V light curve of SNII Linear (continuous line) suggest that the maximum
occurred on JD $\sim 2449675\pm 7$d at a magnitude $V\sim
18.0\pm0.3$}\label{phot_fig}
\end{figure}

The V and R light curves of SN~1994aj are shown in
Fig.~\ref{phot_fig}, including the discovery estimate by Pollas
\shortcite{pol}
and the measurement of Suntzeff \shortcite{sun}.

Patat et al. \shortcite{pat} have shown that SNII can be
separated on the basis of the parameter $\beta_{100}$, which
quantifies the luminosity decline rate in the first 100 days after
maximum light. For SN~1994aj we measure $\beta^{\rm V}_{100}=4.4$
\,mag\, (100d)$^{-1}$ which is typical of the Linear subclass,
very similar for instance to SNe 1980K and 1990K.  The fit of the
light curves of SN~1994aj to the template light curve of SNIIL is
shown. The best fit is obtained if we accept that the maximum occurred
$2-3$ weeks prior to discovery with an uncertainty of $\pm 1$ week and
hereafter we will assume $JD_{\rm max} \sim 2449675$ (Nov 19, 1994) as
reference epoch. Note that this is a conservative choice adopted to reduce
the variance of SN~1994aj photometry compared with that of other SNII.
With this choice we can estimate that the SN reached
a maximum magnitude $V_{\rm max} \sim 18\pm0.3$. However, we are aware that the
above values are subject to considerable uncertainty, since the
maximum was not observed and Type II SNe sometimes show odd
photometric behavior.

After 100-120 days the luminosity decline, both in V and R, suddenly
slowed down: in particular the R light curve from 100 to 535d appears,
considering the photometric error, remarkably linear with
${\gamma}_{\rm R}=0.46$ \,mag~(100d)$^{-1}$.

Qualitatively, this linear decline is reminiscent of ``normal'' SN~IIL
where, however, the late luminosity decline is significantly steeper,
being on the average $\sim 1$\,mag\, (100d)$^{-1}$ \cite{macio2}. 
This last value is in good agreement with the
expectation of the models predicting that, at this phase, the optical
light curve is powered by the thermalization of the $\gamma$-rays and
positrons from the radioactive decay of $^{56}$Co into $^{56}$Fe. If a
fraction of the $\gamma$-rays escapes from the ejecta without being
thermalized the luminosity decline rate may be steeper than the decay
rate (this is thought to occur in SN~Ia). Instead, the case of SN~1994aj,
where the observed decline rate is slower than the radioactive decay
input, is one which requires some additional energy input. In the
following we will argue that this probably results from the
interaction of the ejecta with a dense CSM.

Apart from the moderate galactic reddening, $A_{\rm B}=0.24$ according to
Burstein \& Heiles \shortcite{bh}, there is no evidence of additional
extinction to SN 1994aj.  The SN is projected on an external spiral
arm, close to a very faint HII region, whose narrow emission lines are
seen in the long slit spectra (see next section). The heliocentric
radial velocity at the SN location was obtained by averaging the
measurements of the narrow \Ha~ of different spectra. The result, 
$9530\pm 70 km~s^{-1}$, has been adopted throughout the present
work. After correction for the solar motion relative to the background
radiation ($+304 km~s^{-1}$, RC3) and assuming $H_0=75 km~s^{-1}~Mpc^{-1}$,
we obtain $\mu =35.59$ for the distance modulus of the parent galaxy.

With the above assumptions for extinction and distance, we obtain
$M_{\rm V} = -17.8\pm 0.3$ for the SN absolute magnitude at maximum which,
assuming that the $(B-V)$ color index at
maximum is close to 0, is intermediate between the values for
``regular'' and ``bright'' SNII Linear (respectively $< ~M_{\rm B}~ > =
~-16.8 \pm 0.5$ and $< ~M_{\rm B}~ > = -18.9 \pm 0.6$ \cite{pat}.

\subsection{Spectroscopy} \label{spec}

The journal of the spectroscopic observations reported in
Tab. \ref{spec_tab} gives for each spectrum the date (col.1), the
phase (relative to the adopted maximum, cfr. Sect. \ref{phot})
(col.2), the equipment used (col.3), the exposure time (col.4), the
wavelength range (col.5) and the resolution derived from the average
FWHM of the night-sky lines (col.6). In order to have good wavelength
coverage sometimes exposures obtained with different grisms were merged.
In these cases the cumulative exposure time is reported.

\begin{table}
\caption{Spectroscopic observations of SN 1994aj} \label{spec_tab}
\begin{tabular}{cccccc}
\hline
\hline

     Date     & phase$^*$ & inst.$^{**}$  &  exp. &   range   & resol. \\
              & (days)&       & (min) &   (\AA)   &    (\AA)   \\
\hline
    31/12/94  &  +43  &  2.2  &  60   & 4500-7100 &     11     \\
      3/1/95  &  +46  &  3.6  &  15   & 3700-6900 &     15     \\
      8/1/95  &  +51  &  3.6  & 105   & 3700-9800 &     20     \\
     15/1/95  &  +58  & NTT+E &  40   & 3800-8350 &      8     \\
     16/1/95  &  +59  & NTT+E &  60   & 3800-8350 &      8     \\
     30/1/95  &  +73  & NTT+E & 120   & 3800-8900 &     10     \\
      4/2/95  &  +78  &  2.2  & 120   & 5850-8450 &     11     \\
     28/2/95  & +102  & NTT+E & 120   & 3850-8900 &     10     \\
      3/3/95  & +105  & NTT+E &  60   & 3850-8900 &     10     \\
     29/3/95  & +131  &  3.6  & 120   & 3700-9800 &     16     \\
     30/4/95  & +163  &  3.6  & 225   & 3700-9800 &     20     \\
     29/5/95  & +192  & NTT+E &  60   & 3850-8400 &      8     \\
     14/11/95 & +361  & NTT+E & 120   & 3850-8400 &      8     \\
     17/03/96 & +485  &  3.6  & 180   & 6000-9850 &     18     \\
     11/05/96 & +540  & NTT+E & 120   & 3400-9000 &     14     \\
\hline
\end{tabular}
*  relative to the estimated epoch of maximum JD=2449675

** See note to Table~2 for coding
\end{table}

\begin{figure*}
\psfig{figure=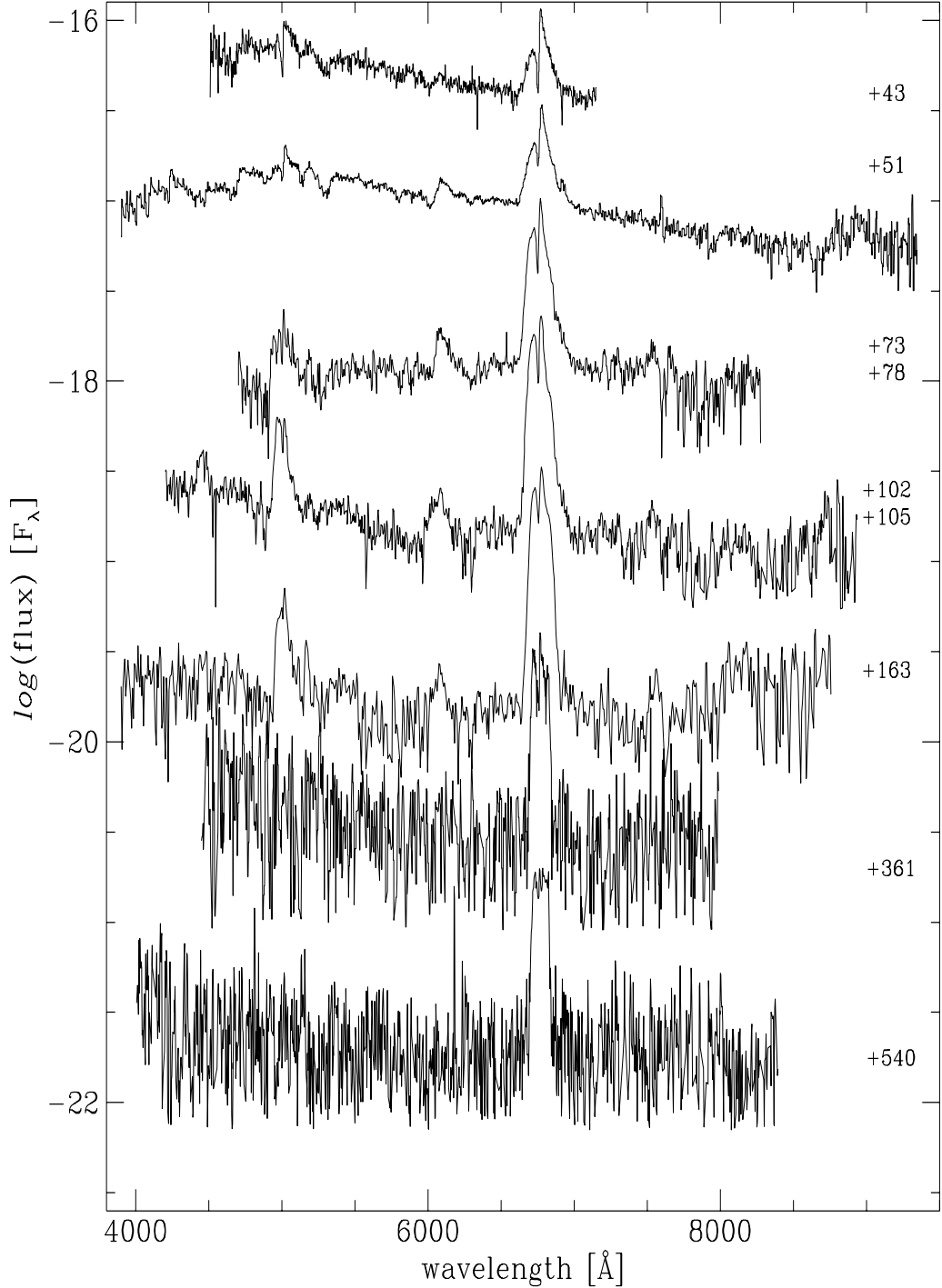,width=16cm,height=20cm}
\caption{Spectral evolution of SN 1994aj. Wavelength is in the
observer rest frame. The ordinate refers to the first spectrum (+43d),
all other spectra are shifted downwards by 0.5 dex with respect to the
one above, with the exception of the last three, for which the shift
is 0.8 dex. If two phases are given for a spectrum, then the average is shown.}
\label{spec_fig}
\end{figure*}

The spectra have been calibrated with He-Ar arcs and flux calibrated
using standard stars from the list of Oke \shortcite{oke}, 
Stone \shortcite{sto}, and Oke \& Gunn \shortcite{okg}.
The absolute flux
calibration of the spectra has been verified against the V and R
photometry. The agreement was generally fair, except for a few cases
in which the spectra were scaled to match the V photometric
measurements.\\
Figure \ref{spec_fig}, which shows a selection of the spectra listed in
Tab. \ref{spec_tab}, illustrates the spectroscopic evolution of SN~1994aj
from phase +43d to +540d.

\begin{figure}
\psfig{figure=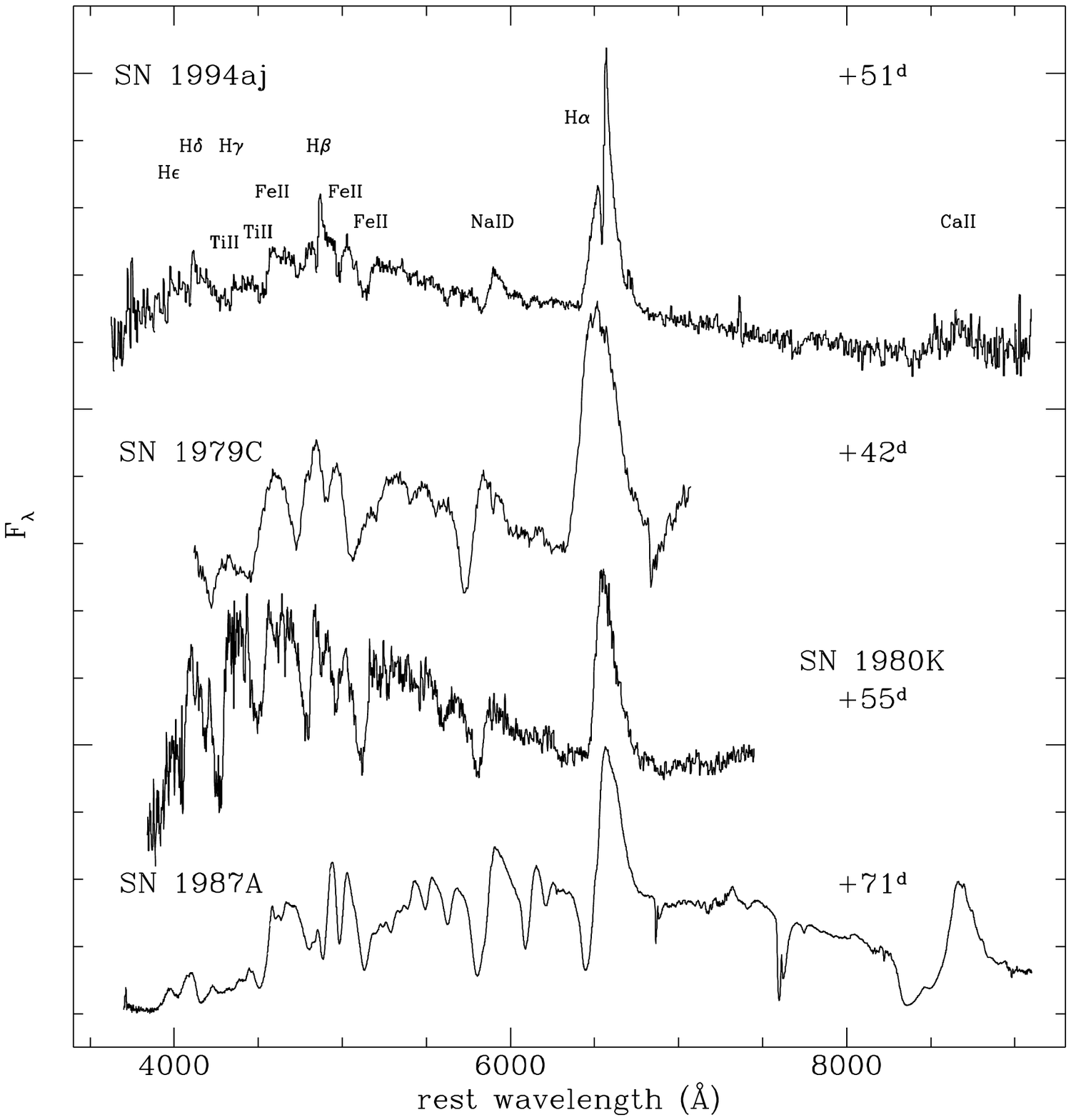,width=9cm,height=10cm}
\caption{Comparison of photospheric spectra of SN 1994aj with the two
Linear SNe 1980K (Benetti, 1993) and 1979C (Branch et al. 1981), and
SN 1987A (ESO database). The spectra have been corrected by the
redshift of the parent galaxy and for galactic
extinction}\label{comp_phot}
\end{figure}

In Fig.~\ref{comp_phot} we compare our best S/N spectrum of SN~1994aj,
obtained at phase 51d, with spectra of the two Linear SNe 1979C
and 1980K, and of SN 1987A of similar phases. The distinctive feature
of SNII, the $H\alpha$ line, dominates the spectra in all cases, but
in SN~1994aj the line has two components, a broad emission component
(FWHM = 170 \AA) on the top of which is a narrow line with a P-Cygni
profile (FWHM = 25
\AA~ for the emission). Whereas the profile of the $H\alpha$ line will
be discussed in more detail later, let us note that there is no
evidence of broad P-Cygni absorption although the line is asymmetric
with a more extended red wing. This is similar to the case of
SN~1979C and other Linear SN~II and has been attributed to the fact
that the H$\alpha$ feature originates only partially through
redistribution of the continuum light from the photosphere; there
is also a net H$\alpha$ emission due to collisional excitation of H
\cite{branch}. The effect is usually stronger in
H$\alpha$ because of the lower energy required compared, for instance,
to H$\beta$ and owing to the large optical depth of this line. 
% This, apparently, is not true for SN~1994aj, where neither H$\beta$
% nor H$\gamma$, and possible H$\delta$, show broad P-Cyg absorption at
% this epoch, but a profile similar to that of H$\alpha$.

The spectrum of SN1994aj has a continuum temperature of about 6500 K,
slightly higher than that of SN~1987A at the same phase, but similar
to that of SN~1979C.

As is the case in SN~1979C, other lines present in the spectrum of
SN~1994aj show the broad P-Cygni absorption, in particular the NaID,
CaII infrared triplet and possibly the two FeII lines of multiplet 42,
$\lambda \lambda$ 5018\AA~ and 5169\AA, all usually quite strong in the
spectrum of SNII at this phase.\\
It should be noted that the expansion velocity derived from the
minimum of the P-Cygni absorptions, $\sim 2500 km/s$, is smaller than
that of SN~1979C, $\sim 8000 km/s$, but instead very close to that
measured for other Linear SN~II, e.g. 1980K and 1990K, and for
SN~1987A around this phase \cite{hanu}.

At a phase 100-200d after maximum, the emission lines grow in
strength relative to the continuum (see Fig.~\ref{late}). The spectrum
appears dominated by the Balmer line of H and the NaID lines is also
conspicuous. At these phases, H$\beta$ shows a hint of broad P-Cyg
absorption, from which an expansion velocity of $\sim 6500$ km/s is
derived. The FeII lines of multiplet 42 and 49 and the CaII] 7291-7324
\AA~ lines are present, although the latter is weaker than is normally
observed in SNII. The [OI] 6300-6364 \AA~ lines, are completely absent.

One year after maximum only a strong H$\alpha$ emission is detected in
the spectrum (see Fig.~\ref{comp_neb}). The complete absence of
forbidden line emission, in particular [OI], [OII] and CaII] in the
late spectra of SN~1994aj should be noted: at phase 485d we can put a
lower limit to the ratio of the H$\alpha$ to CaII] 7291-7324 \AA~
doublet $\log [F(H\alpha)/F(CaII)] \ge 1.2$ instead of typical value
of 0 to $-0.5$ (e.g. Cappellaro et al. 1995b). A possible explanation
of this characteristic is that the density in the line emitting region
is much higher than the typical value for SNII at this phase, that is
$n_{\rm e} \sim 10^8 cm^{-3}$ (Branch at al. 1981; Spyromilio \& Pinto 1991).
The absence of oxygen emission lines may also point to an
underabundance of this element in the SN envelope.

\subsection {The profile of \Ha\ emission} \label{ha}

In Fig.~\ref{Ha_profile} the spectra of SN~1994aj are expanded to show
the evolution of the complex profile of the H$\alpha$ line. In Table 4
are reported the main \Ha~ parameters as measured with the ALICE
package in MIDAS, which allows for multiple Gaussian fitting of
complex line profiles.

The \Ha~ profile at 51d consists of the broad component
($FWHM = 7500 km/s$) which is clearly asymmetric with the red wing
zero intensity velocity of $\sim 10000$ km/s (see Tab. \ref{line}),
significantly larger than the corresponding velocity of the blue wing,
6500 km/s.  On top of this feature is a relatively narrow P-Cygni
feature whose emission component is centered at the same wavelength as
the peak of the broad component. The blue wing of the absorption
indicates a maximum velocity of $\sim 2000$ km/s (see Tab. \ref{line})
whereas the minimum of the trough corresponds to $\sim 900$ km/s.

It is important to note that the narrow P-Cygni feature appears also
on the other Balmer lines visible in the spectra, namely H$\beta$ and
H$\delta$. Although, the S/N and resolution are not good enough to
analyze in detail the profiles of these latter lines, the positions of
the absorption minima are consistent with the same expansion velocity
as the H$\alpha$ line. Concerning the narrow P-Cygni feature of
\Ha\ we note that for the first 200d, its profile remains symmetric
(the intensity of the absorption is similar to that of the emission,
see Tab. \ref{line}) suggesting that it is produced by scattering of
the photons coming from the SN photosphere, by the hydrogen atoms of
the CSM. The intensities gradually decrease and in the latest
spectra, corresponding to 1.3 and 1.5 yr after maximum, the narrow
absorption feature is barely visible. The velocity corresponding to
the minimum of the absorption maintains an almost constant value of
about 900 km/s, for the first 100d then decreases to about 500 km/s at
200d.

\begin{table*}
\caption{\Ha\ Parameters} \label{line}
\begin{tabular}{cccccccc}
\hline
\hline

 Phase & Total   & $^{*}\lambda_{blue-wing}$ & $^{*}\lambda_{red-wing}$ & Flux narrow & Em/Ab  & $^{*}\lambda_{blue-wing}$ \\
       & flux    & broad comp. & broad comp. & comp. (Em)  & (narrow) & narrow abs.      \\
 (days)& ($\times 10^{-16} erg~s^{-1} cm^{-2}$) & (\AA) & (\AA) & ($\times 10^{-16} erg~s^{-1} cm^{-2}$) & & (\AA) \\
\hline
  +43  & 70.6 & 6594 & 6962 & 7.8 & 0.88 & 6704 \\
  +46  & 90.0 &	6604 &  --  & 7.3 & 1.07 & 6712 \\
  +51  & 83.9 &	6617 & 7010 & 6.2 & 1.19 & 6700 \\
  +59  & 94.7 &	6616 & 7000 & 5.9 & 1.02 & 6699 \\
  +73  & 97.3 &	6620 & 7004 & 5.5 & 0.95 & 6711 \\
  +78  & 76.4 &	6626 & 7001 & 4.5 & 1.02 & 6720 \\
 +102  & 68.7 & 6630 & 6990 & 3.8 & 1.09 & 6726 \\
 +105  & 76.0 &	6636 & 6996 & 4.0 & 0.98 & 6728 \\
 +131  & 71.0 &	6633 & 6957 & 3.5 & 0.97 & 6728 \\
 +163  & 70.9 &	6640 & 6957 & 5.1 & 1.13 & 6731 \\
 +192  & 83.0 &	6653 & 6887 & 4.4 & 1.02 & 6726 \\
 +361  & 40.0 &	6681 & 6850 & 3.8 & 4.75: & 6718 \\
 +485  & 11.0 &	6679 & 6856 & 1.1 &  --  &  --  \\
 +540  & 14.5 &	6686 & 6846 &  -- &  --  & --   \\
\hline		
\end{tabular}

$*$ Wavelength in the observer rest frame\\
\end{table*}

With time, the broad component became progressively more symmetric
until in the spectra of one year or later it shows a roughly rectangular
profile with a FWZI $\sim$ 6000 km/s (See Tab. \ref{line}), which is
consistent with emission from a shell of gas whose thickness is much less 
than the radius of the shell. Actually
in the latest spectra there is a hint of a double peak structure for
the broad component which could indicate that the interaction between
the SN ejecta and CSM is asymmetric.

Narrow P-Cygni features have been noted in the early spectra of two
other SNII Linear, namely 1979C and 1984E. In SN~1979C \cite{branch}
the feature is present as the spectra
evolve, but its relative strength is never as strong as in SN1994aj,
whereas in SN~1984E the feature is quite strong only near
discovery, and has already disappeared one month later \cite{hb},
indicating that the CSM was confined close to the
progenitor. Moreover, in SN 1984E the P-Cygni emission component is
significantly stronger than absorption \cite{dop},
whereas in SN~1994aj they have comparable strength.

For all these SNe the narrow P-Cygni features can be attributed to a
super-wind episode suffered by the SN precursor some time before
explosion. From the observations presented here we can derive some
constraints for the wind characteristics:

\begin{enumerate}
\item the wind has an inner boundary, or at least a density peak, 
at velocity $v \sim 500$ km/s and extends with a negative density
gradient up to a maximum velocity of $~\sim 2000$ km. We suggest that
at early phases the minimum of the absorption corresponds to a
velocity higher than that of the density peak both because of large
optical depth of the line and because of a shallow density gradient of
the circumstellar wind as demonstrated by Jeffery \& Branch \shortcite{jb}.

\begin{figure}
\psfig{figure=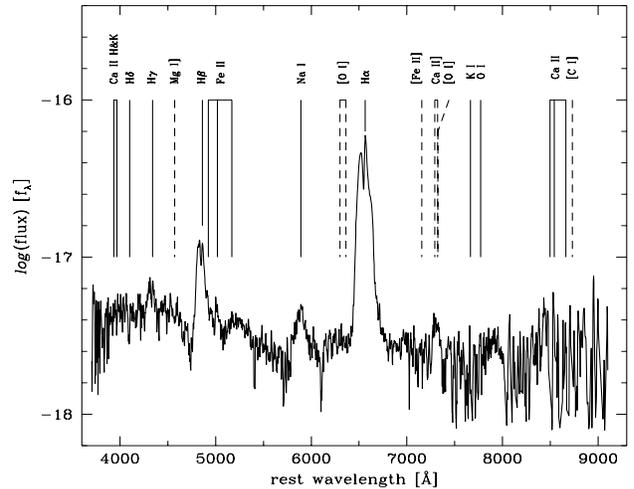,width=9cm,height=8cm}
\caption{Average of the four spectra between phase +102d and +163d
with the line identifications. In the figure some lines normally
present in type II supernovae at this phase, but not in SN 1994aj are
also marked with dashed lines.}\label{late}
\end{figure}

\item It is expected that the burst of UV radiation that occurs when
the shock emerges at the surface is strong enough to completely ionize
the circumstellar matter surrounding the SN (Fransson \& Lundquist
1989; Nadyozhin 1994).  Since we observe narrow P-Cyg absorption in
the HI lines a significant fraction of the H must have recombined on a
time scale shorter than a few weeks. This requires that the density of
the super-wind at the outer edge of the emitting region must be
$n_{\rm e} > 10^5 cm^{-3}$.  Of course this does not exclude that a
low density wind is present at larger radii, since this wind will not
have strong lines in the optical region.

\item   As we will see in the next section there is evidence of 
interaction between the ejecta and the circumstellar matter starting
100d after maximum.  Since the maximum expansion velocity of the
ejecta was seen to be $\sim 10000 km/s$ we derive an inner radius for
the wind $r_{\rm i} \sim 10^{16}$ cm.  Given the wind velocity, this means
that the superwind episode terminated only a few years, $\sim 5-10$, before
explosion.  On the other hand the interaction is still strong up to
the last available spectrum (540d) which indicates that the wind has
not yet been swept up completely. This means that the outer edge of
the wind has a radius $r_{\rm e} > 5 \times 10^{16}$ cm. Assuming a steady
superwind (with a constant terminal outflow speed $V_\infty$ and
a constant mass loss rate $\dot{M}$), i.e.  a density
varying with $r^{-2}$, we can obtain a lower limit for the mass of the
super-wind:

$$ M \ge 4\pi r_{\rm e}^3 m_{\rm H} n_{\rm e}(r_{\rm e}) (1-\frac{r_{\rm i}}{r_{\rm e}}) \sim 0.1 M_{\odot}$$

\begin{figure}
\psfig{figure=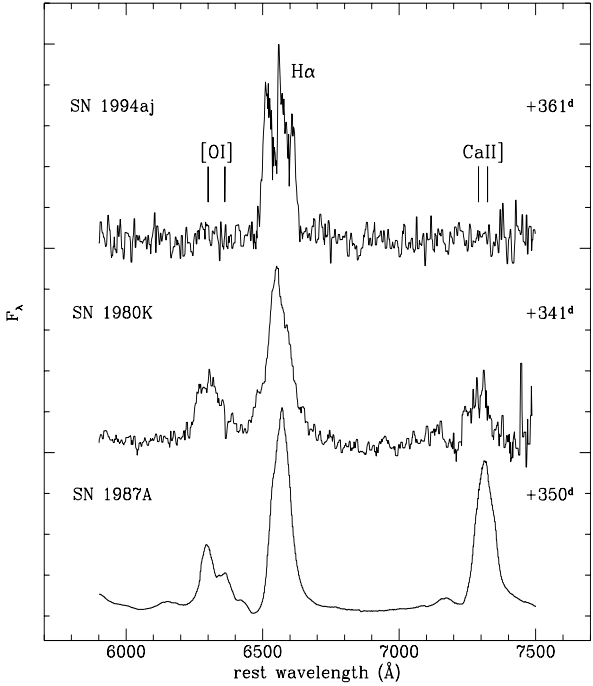,width=9cm,height=10cm}
\caption{Comparison of the late time spectra of SN 1994aj, SN 1980K
(Uomoto \& Kirshner, 1986), SN 1987A (ESO database). The spectra have
been corrected by the redshift of the parent galaxy and for
extinction}\label{comp_neb}
\end{figure}

%\item The rectangular profile is consistent with emission from a shell of gas
%whose thickness is much less than the radius of the shell.
%
\end{enumerate}

% The evolution of the \Ha~ profile in the spectra of SN 1994aj is shown
% in Fig. \ref{Ha_profile}. 

\begin{figure}
\psfig{figure=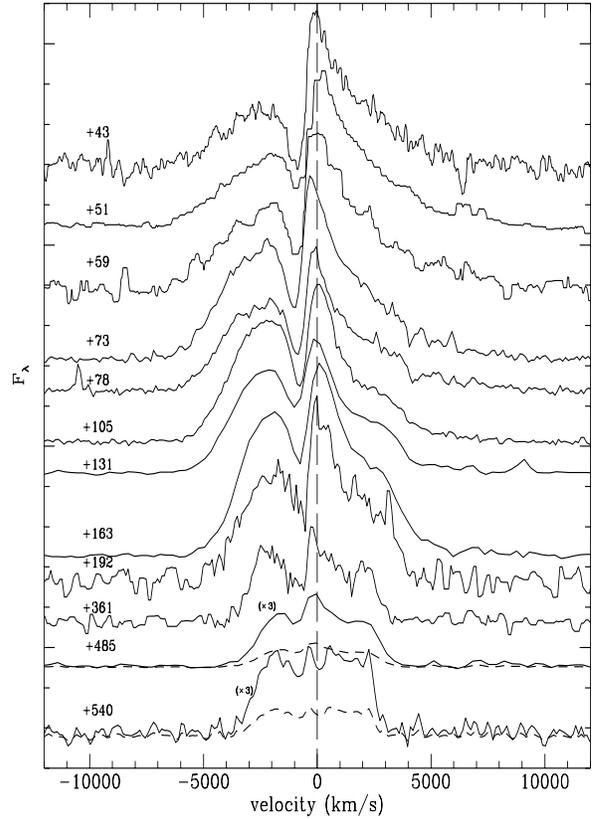,width=9cm,height=12cm}
\caption{Evolution of the \Ha~ profile in SN 1994aj. The long dashed
line marks the galaxy rest frame position of \Ha~ transition. For an
easier interpretation the last two spectra have been multiplied by a
factor of 3 (the original spectra are plotted with a short dashed
line).}\label{Ha_profile}
\end{figure}

\subsection{Flux} \label{fluxes}

\begin{figure}
\psfig{figure=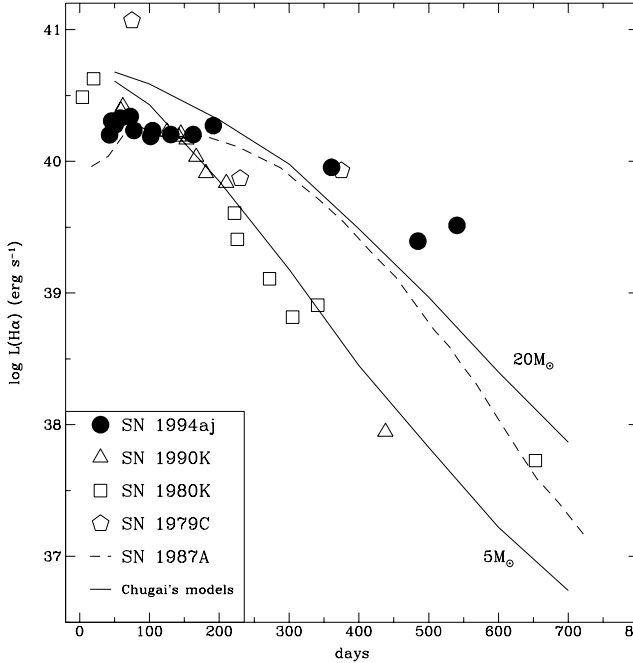,width=10cm,height=10cm} \caption{Evolution
of the \Ha~ luminosity of SN 1994aj compared with the same data for
SNe 1980K (Uomoto \& Kirshner, 1986), 1990K (Cappellaro et al. 1995b),
1979C (Branch et al. 1981; Chevalier \& Fransson 1985) and 
1987A (Danziger et al. 1991). Also
reported are the Chugai's models (Chugai 1991).}\label{Halpha-flux}
\end{figure}

In Fig.~\ref{Halpha-flux} we compare the absolute \Ha~ light curve of
SN~1994aj (the observed \Ha\ fluxes are also given in Tab. \ref{line})
with that of the Linear Type SN~1990K, SN~1980K and SN~1979C, and of
SN~1987A. There is agreement amongst this sample up to $\sim 150d$
past maximum, but at later phases SN~1994aj declines at a much slower
rate than the others except for SN~1979C. In the figure are also shown
the predicted behavior of models in which the late input energy
derives only from radioactive decay of Co to Fe \cite{chu}. Two models
with different ejecta mass, namely 5 and 20 M$_{\odot}$, are shown.

The differences between SNe 1980K, 1990K and 1987A can be understood
in terms of different ejecta masses. Linear SNII are fitted by a model
with a small mass of the ejecta ($\le $5 M$_{\odot}$) which is consistent with
the interpretation of the early light curve and with evidence from the
spectra \cite{capp}. A much higher mass of the
ejecta, 10-15 M$_{\odot}$, is instead required to fit the late H$\alpha$
emission of SN~1987A which is consistent with the early light curve.

The early V and R light curves and spectra of SN~1994aj are very
similar to those of other Linear SNII, in particular 1980K and
1990K. This implies that the ejecta mass is similar, even though the
late H$\alpha$ luminosity is higher than for SN~1987A. Since there is
evidence of a dense circumstellar medium around SN~1994aj it is
natural to attribute the excess of luminosity to the conversion of
kinetic energy into radiation when the ejecta interacts with the
circumstellar medium. Evidence of a similar phenomenon are now
available for a number of SNII, SN 1988Z \cite{cd}, which shows the
most powerful and long lasting emission in H$\alpha$ ever seen in a
supernova (Turatto et al. 1993, 1997); SN 1978K \cite{cdd} and in
particular two more recent objects which are now under intensive
studies: SN 1996L \cite{ben2} and SN 1996al \cite{bn}.  We can
estimate the amount of energy released in the interaction, by
subtracting, from the observed luminosity, the contribution due to the
radioactive decay of Co in Fe which we assume is well represented by
the 5 M$_{\odot}$ model of Chugai (the result does not change
significantly if we assume a mass of 10 M$_{\odot}$). We found that the
excess of luminosity is more or less constant with phase and of the
order of $4\times10^{39}$ erg/s, for phases $> 200d$.

We can give a rough estimates of the density and mass loss rate for
the wind by using the equation

$$ L(H\alpha) = \frac{1}{2} \psi \frac{\dot{M}}{u_{\rm w}} v^3_{\rm sn}$$

\noindent
where $\psi$ is the efficiency for converting kinetic energy into
H$\alpha$ radiation, $\dot{M}$ is the mass loss rate, $u_{\rm w}$ the wind
velocity, $v_{\rm SN}$ the velocity at the shell-wind contact
discontinuity. $\psi$ is not well known, but of the order of $0.1$
\cite{chu1}.\\ 
By adopting for $v_{\rm sn}$ the mean value of the half-width at zero
intensity of the H$\alpha$ line for the last three spectra, $\sim
3700$ km/s we derive $\dot{M}/u_{\rm w} \sim 1.7 \times 10^{15}$ g/cm.
Similar values are reported by Chugai for two linear SNe 1970G and SN
1980K, and this is four times smaller than the value derived for SN
1979C. In all these cases however the interaction began a few years after
the explosion.

For a normal stellar wind with $u_{\rm w} = 10$ km/s this would imply a mass
loss rate $\dot{M} \sim 10^{-5}$ M$_{\odot}$/yr. However, most likely
the interaction is occurring with the material of the circumstellar shell
which gave
rise to the narrow P-Cyg feature in the early phases. If this is true, then 
the velocity is of the order of $10^{3}$
km/s, and the mass loss rate should therefore be $\dot{M} \sim
10^{-3}$ M$_{\odot}$/yr.

\section{Conclusions}

From the observations presented in this paper we conclude that SN 1994aj was
a type II SN with linear light curve.\\ 
It is generally believed \cite{bb,cd} that Linear SNII
result from the explosion of
stars with main sequence masses $\sim 8-20 M_{\odot}$ which have lost
most of their envelope via stellar winds. At the time of explosion the
mass of the progenitor is of the order of $3-5 M_{\odot}$ and the star is
surrounded by several solar masses of circumstellar matter. The
explosion produces about $0.1 M_{\odot}$ Ni which powers the light curve via
radioactive decay.

While for SNII-L the low mass of the ejecta causes the rapid
luminosity decline of the light curve, the detailed SN output depends
on the distribution of the circumstellar material at the time of
explosion.\\ 
The observations of SN 1994aj may
then correspond to the following scenario at the time of the explosion.

A dense ($\dot{M}/u_{\rm w} \sim 10^{15}$ g/cm) wind, resulting from the
late evolution of the progenitor, is confined to a
relatively short distance from the star ($R_{\rm out} = few
10^{16}$ cm). The circumstellar material is
completely ionized by the burst of hard radiation, but because of the
high density it rapidly recombines. The re-radiation of
the UV photons created at shock break-out may give rise to a very high optical
luminosity at maximum observed for instance in the case of SN~1979C
(Branch et al. 1981; SN~1994aj was not observed near maximum). While cooling,
the shell becomes visible through Balmer line scattering, for which the
P-Cygni profiles indicate a large velocity field (from a few $10^2$
km/s to 2000 km/s for SN~1994aj and similar values for SN~1984E and
1979C). In the case of SN~1979C and for some weeks after explosion, UV
emission lines from highly ionized atoms (e.g. NV, SiIV, CIV) have
been observed with even larger maximum velocity (up to 4000
km/s). These lines probably originate in the outer regions of lower
density \cite{branch}.

Some time after explosion the ejecta impacts upon the shell and its
kinetic energy is converted into radiation, slowing down
the decline of \Ha~(and R) luminosity. In the case of SN~1994aj this
begins about 100-150 days after maximum and is still active at 540
days.  The observed excess of luminosity is consistent with the wind
parameters (density and velocity) reported above if we accept an episode
of very strong mass loss rate ($\dot{M} = 10^{-3} M_{\odot}/yr$)
which terminated shortly before explosion ($\sim 5-10$ yr).

There is evidence (linear light curves and the absence of [OI]
6300-6364 \AA\ emission lines in the nebular spectra) for a lower mass
of the SN 1994aj ejecta ($< 5 M_{\odot}$), which would indicate an origin
of this supernova from a main sequence star with a mass about $8 - 10
M_{\odot}$ \cite{cd}.\\
Chugai \& Danziger suggest a similar mass range for the progenitor of
the peculiar type IIn SN~1988Z, for which they find that less than 1
M$_{\odot}$ was ejected in the explosion. The photometric and
spectroscopic differences between SN~1988Z and SN~1994aj might then be
explained by differences in the mass loss history along the
progenitor's evolution, and differences in the density and distribution
of the ISM in which the two progenitors were embedded at the time of
their evolution and explosion.

Given the CSM characteristics of SN 1994aj, an expected radio luminosity
between $10^{25}-10^{26}
erg\,s^{-1} Hz^{-1}$ at 6cm is derived (Chevalier 1982; Van Dyk 1996).
Such luminosity, given the
distance of SN 1994aj, would be beyond the detection limit of current
radio telescopes. The difficult detection of another Type IIL
supernova, SN~1986E, an object much closer than SN~1994aj, at 0.304
mJy by Montes et al. \shortcite{montes} reinforces this conclusion.

The X-ray luminosity depends on the square of ($\dot{M}/u_{\rm w}$)
\cite{chu1}, and for SN 1994aj this is $L_{\rm X} \sim 10^{39}$ erg/s,
placing the supernova (given its distance) far below the detection
limit of existing X-ray instruments. Coronal lines have not been
detected in the optical spectra.

Another possible scenario that may explain some of the SNII-properties
is given by Nomoto et al. \shortcite{no}. In this scenario the Fe core
collapse of massive ($\ge 10 M_{\odot}$) stars in close binary systems is
invoked. The variety of Type II supernovae (II-L, IIb and IIn, and
conceivably also Ib and Ic supernovae) are explained with the merging
of the two stars in a non-conservative mass transfer scenario. During
the merging of the stars a significant fraction of the common envelope
is lost due to frictional heating. The difference in the SN types
originates from the difference in the mass of the H-rich envelope. In
this scenario the CSM would consist of the material ejected during the
spiraling-in and the RSG wind material, where the structure of the CSM
originated from spiraling-in, is likely to be asymmetric. This could
naturally explain asymmetries in the ejecta - CSM interaction such as
the one possibly seen from the evolution of the \Ha\ profile of SN
1994aj (see Sect.~\ref{ha}). Intensive mass loss such as that seen in
SN 1994aj might be better explained in this last ``dynamic''
scenario.\\ 
In any case, any model interpreting SNII-L properties should be able
to account for the strong mass loss experienced by these SNe just
before explosion.

\bigskip
 
\noindent
{\bf ACKNOWLEDGMENTS} We thank L. Koesterke and R. Gilmozzi
for giving us part of their telescope time.

\end{document}